\documentclass[fleqn,12pt,twoside]{article}

\usepackage{graphicx}
\usepackage[headings]{espcrc1}
\usepackage{bm}
\usepackage{epsfig}
\usepackage{amsmath}
\usepackage{amssymb} 
\newcommand{\R}{\mathbb R}

\title{How complex a complex network of equal nodes can be?}
\author{M.  S.  Baptista\address[CMUP]{Centro de Matem\'atica da Universidade do Porto, Rua do Campo
Alegre 687, 4169-007 Porto, Portugal}
\thanks{previous address: Max-Planck-Institut f\"ur Physik komplexer Systeme, 
N\"othnitzerstr. 38, D-01187 Dresden, Germany,
future address: Institute for Complex
Systems and Mathematical Biology, King's College, University of
Aberdeen, AB24 3UE Aberdeen, United Kingdom}, 
F. Moukam Kakmeni\address[DFFB]{Department of Physics, Faculty of
  Science, University of Buea, P. O. Box 63 Buea, Cameroon}\thanks{previous address: Max-Planck-Institut f\"ur Physik komplexer Systeme,
N\"othnitzerstr. 38, D-01187 Dresden, Germany}, 
Gianluigi Del Magno\address[MPIPKS]{Max-Planck-Institut f\"ur Physik komplexer Systeme,
N\"othnitzerstr. 38, D-01187 Dresden, Germany}
, M. S. Hussein\address[IFUSP]{Institute of Physics, University of S\~ao Paulo,
Rua do Mat\~ao, Travessa R, 187, 05508-090 SP, Brasil}}

\runtitle{How complex a complex network of equal nodes can be?}
\runauthor{M. S. Baptista}

\begin{document}
\maketitle

\begin{abstract}
  Positive Lyapunov exponents measure the asymptotic exponential
  divergence of nearby trajectories of a dynamical system. Not only
  they quantify how chaotic a dynamical system is, but since their sum
  is an upper bound for the entropy by the Ruelle inequality, they
  also provide a convenient way to quantify the complexity of an
  active network. We present numerical evidences that for a large
  class of active networks, the sum of the positive Lyapunov exponents
  is bounded by the sum of the positive Lyapunov exponents of the
  corresponding synchronization manifold, the last quantity being in
  principle easier to compute than the latter. This fact is a
  consequence of the property that for an active network considered
  here, the amount of information produced is more affected by the
  interactions between the nodes than by the topology of the network.
  Using the inequality described above, we explain how to predict the
  behavior of a large active network only knowing the information
  provided by an active network consisting of two coupled nodes.
\end{abstract}


\section{Introduction}

The relation between topology and function in active networks,
networks composed by nodes described by some intrinsic deterministic
dynamics, is a fundamental question whose answer may help understand
the collective behavior \cite{strogatz} of a variety of complex
systems ranging from particle-like chemical waves \cite{showalter},
light propagation in dieletric structures \cite{amann}, neural
networks \cite{jacob} and metabolic networks \cite{steuer}.

The work of Kuramoto \cite{kuramoto} and the works of Pecora and
collaborators \cite{pecora,pecora1} laid the foundations of a
theoretical framework for studying the relation between topology and
function in active networks. In particular, the latter opened up a new
way to study the onset of complete synchronization in active networks
\cite{stefano,juergen,hasler} composed of equal node dynamics.

At the present moment, it is important to understand from a
theoretical perspective the relation between the structure of a
network (topology) and the behavior of it (function) in active
networks whose nodes are not only far away from complete
synchronization (desynchronous) but also nodes that interact among
themselves simultaneously by linear and nonlinear means.

In this work, we conjecture that an upper (or lower) bound for the sum
of the Lyapunov exponents of an active network with some special
properties \cite{coment_networks} and an arbitrary size, formed by
nodes possessing equal dynamics, can be analytically calculated by
only using information coming from the behavior of two coupled nodes.
We recall that by the Ruelle Formula \cite{comentario}, the sum of the
positive Lyapunov exponents is an upper bound for the entropy. Hence,
the sum of the positive Lyapunov exponents represent a convenient way
to quantify the behavior of the network and therefore to measure how
complex a network is. 

\begin{figure}[!h]
  \centerline{\hbox{\psfig{file=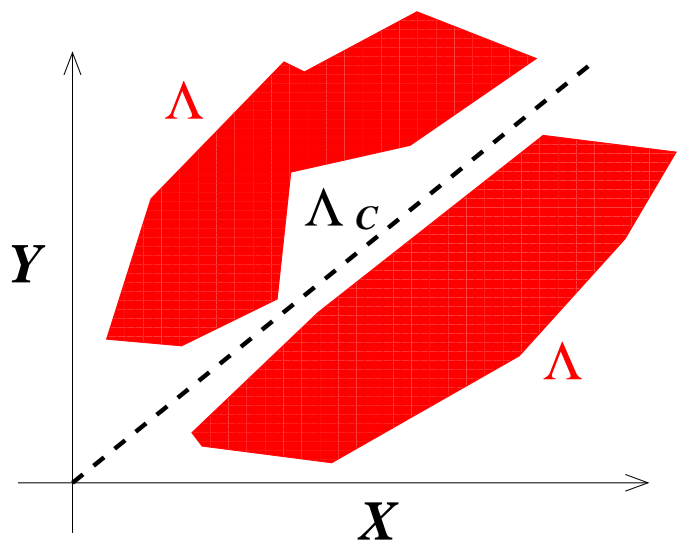,height=8.0cm,width=8.0cm,
        viewport=0 0 500 510}}} \caption{[Color online] Illustration
    of the two most relevant types of solutions we expect to find in
    the networks here considered. A synchronous solution whose
    trajectory is represented by the black dashed line, which lies on
    the synchronization manifold, and the desynchronous solution whose
    trajectory is represented by the red filled regions. The sum of
    the positive Lyapunov exponents of the synchronous solution is
    denoted by $\Lambda_C$ and the sum of the positive Lyapunov
    exponents of the desynchronous solution is denoted by $\Lambda$.}
\label{ilustra_conjectura}
\end{figure}

To describe our conjecture, we first introduce some concepts and
ideas, illustrated by Fig. \ref{ilustra_conjectura}. This figure
represents the trajectory of two nodes $X$ and $Y$ of a large
network. The networks considered here admit a synchronous solution
[see Eq.  (\ref{element_dynamics})] and a desynchronous one.  The
position where this synchronous solution lies is pictorially
represented by the dashed black line that represents a projection of
the synchronization manifold of the network.  The desynchronous
solution is represented by the filled red regions localed off the
diagonal. This solution represents a chaotic desynchronous trajectory.

If the synchronous solution is unstable, initial conditions close to
the synchronization manifold leave its neighborhood, eventually
arriving at a desynchronous (stable) solution, a chaotic attractor.
If the synchronous solution is stable, it is to be expected that
complete synchronization takes place, when all nodes have equal
trajectories.

The Lyapunov exponents of the desynchronous solutions (a chaotic
attractor) are calculated from Eq.  (\ref{variational0}), and the sum
of the positive ones is denoted by $\Lambda$. The Lyapunov exponents
of the synchronous solution are refered to as conditional Lyapunov
exponents, and the sum of the positive ones is denoted by $\Lambda_C$.
\cite{comentario_LE}.

Roughly speaking, our conjecture states that if for two ($N=2$)
coupled nodes with equal dynamics and coupling strengths, the quantity
$\Lambda$ is greater (smaller) than $\Lambda_C$, then this inequality
remains valid for $N > 2$ coupled nodes (with the same dynamics) with
coupling strengths obtained by properly rescaling.

Accordingly, given an interval for each coupling strength, the
collection of all networks considered here can be classified in two
classes : The class LOWER for which $\Lambda \geq \Lambda_C$
($\Lambda_C$ is a lower bound for $\Lambda$) and the class UPPER for
which $\Lambda \leq \Lambda_C$ ($\Lambda_C$ is an upper bound for
$\Lambda$).  While for the first class, a node forces another not to
do what it is doing, inducing the nodes to stay out of synchrony, in
the second class a node forces another to do what it is doing,
inducing all the nodes to become synchronous.

Naturally, if the nodes in the network becomes completely synchronous,
then the synchronous solution becomes stable and $\Lambda =
\Lambda_C$.

It is often considered that the complexity of a network can be
quantified by typical characteristics as the average degree, the
network's connecting topology, the minimal and maximal degree, the
average or minimal path length connecting two nodes, and others. But
these characteristics are a measure of the structure of the network
and not of the behavior of it. In this work, at least for the class of
networks considered here, we can state that these active networks
behave in only two ways, regardless the many characteristics that
quantify the network's structure: the behaviors UPPER and LOWER. In
other words, if nodes of an active network with equal nodes interact
by a coupling function that induces an LOWER (or UPPER) character,
this character will not be modified by the use of other connecting
topologies.


To justify our conjecture, we use complex networks of linear and
nonlinear maps coupled by linear terms, and neural networks of highly
non-linear neurons (Hindmarsh-Rose (HR) neurons \cite{HR}) connected
simultaneously by linear couplings (electrical synapses) and
non-linear couplings (chemical synapses).

We finally discuss how our conjecture can be used to predict whether a
network formed by nodes that when isolated are chaotic (periodic) will
maintain such a chaotic behavior, then predicting how complex larger
networks can be.

\section{Active networks}

Consider an active network formed by $N>0$ equal nodes $\mathbf{x_i}
\in
\R^{d}$ with $ d>2 $. The network
is described by
\begin{equation}\label{element_dynamics}
\dot{\mathbf{x_i}}= \mathbf{F(x_i)} + \sigma
\sum_{j=1}^{N}\mathcal{G}_{ij}\mathbf{H(x_j)} +
g\sum_{j=1}^{N}\mathcal{C}_{ij}\mathbf{S(x_i,x_j)},
\end{equation} 
where $ g \in \R $ and $ \sigma>0 $,
$\mathcal{G}=\{\mathcal{G}_{ij}\}$ is a Laplacian matrix ($\sum_j
\mathcal{G}_{ij} = 0$) describing the way nodes are linearly coupled,
$\mathcal{C}=\{\mathcal{C}_{ij}\} $ is the the adjacent matrix
representing the way the nodes are connected by linear and non-linear
function, and $\mathbf{H}:\R^{d} \to \R^{d} $ and $\mathbf{S}:\R^{d}
\times \R^{d} \to \R^{d} $ are arbitrary differentiable
transformations. We also assume that $ \mathcal{G} $ and $ \mathcal{C}
$ commute.

A solution of (\ref{element_dynamics}) is called synchronous if $
\mathbf{x_1}(t) = \cdots = \mathbf{x_N}(t) $. To guarantee the
existence of such solutions, we assume that every node of the network
receives the same number $k$ of incoming connections. In other words,
we require that $\sum_j \mathcal{C}_{ij} = k$ for any $i$. It is easy
to see that this condition not only guarantees the existence of
synchronous solution, but also implies that the $ d $-dimensional
linear subspace $ \mathcal{S} = \{\mathbf{x_1 = x_2 = \ldots = x_N}\}$
is invariant. The set $ \mathcal{S} $ is called synchronization
manifold. Note that a synchronous solution $
\mathbf{x}_{i}(t)=\mathbf{x}(t) $ for $ i=1,\ldots,N $ satisfies the
following ordinary differential equation
\begin{equation}
	\label{synchronous}
\dot{\mathbf{x}}= F(\mathbf{x}) + g k \mathbf{S(x,x)}.
\end{equation}


The way small perturbations $\mathbf{\delta x}_1,\mathbf{\delta
x}_2,\ldots,\mathbf{\delta x}_N$ propagate in the network is described
by the variational equations \cite{pecora} associated to
(\ref{element_dynamics})
\begin{eqnarray}\label{variational0} 
\dot{\mathbf{\delta x}}_i &=& DF(\mathbf{x}_i)\mathbf{\delta x}_i + \sigma \sum_{j=1}^N
 \mathcal{G}_{ij} D {\mathbf{H}}(\mathbf{x}_j)\mathbf{\delta x}_j + \\
 && g\sum_{j=1}^N \mathcal{C}_{ij} D_{1}
 \mathbf{S}(\mathbf{x}_i,\mathbf{x}_j) \mathbf{\delta x}_i + g\sum_{j
 \neq i} \mathcal{C}_{ij} D_{2} \mathbf{S}(\mathbf{x}_i,\mathbf{x}_j)
 \mathbf{\delta x}_j,\nonumber
\end{eqnarray}
where $ D_{1} S(x,y) $ and $ D_{2} S(x,y) $ denote the differential of
$ S(x,y) $ with respect to $ x $ and $ y $, respectively. From
(\ref{variational0}), we can calculate the Lyapunov exponents of every
solution of (\ref{element_dynamics}). The network is assumed to be
ergodic, and so the Lyapunov exponents $ \lambda_{1} \le \lambda_{2}
\le \cdots \le \lambda_{m} $ for $ m=1,\ldots,Nd $ are constant almost
everywhere, and can be obtained by typical initial conditions. The
$Nd$ Lyapunov exponents of the synchronous solutions are called
conditional Lyapunov exponents. We also assume that the dynamics
restricted to the synchronization manifold $ \mathcal{S} $ is ergodic.
Hence, also the conditional Lyapunov exponents along synchronous
solutions are constant almost everywhere on $ \mathcal{S} $. The
ergodic invariant measure of (\ref{element_dynamics}) and that of the
dynamics restricted to $ \mathcal{S} $ (not necessarely the same) are
assumed to be unique (singular) and different than a point
(non-atomic).

\section{Conjecture}

Here, we describe our proposed conjecture in a more friendly way. For
a more rigorous presentation of it, one should read the Appendix
\ref{conjecture}.

Let $ \mathbf{H,S,\mathcal{G},\mathcal{C}},\sigma,g,N$ as in
(\ref{element_dynamics}) to be the parameters which define the active
network.  $\mathbf{H}$ represents the function under which the nodes
connect among themselves in a linear fashion, $\mathbf{S}$ the
function under which the nodes connect among themselves in a
non-linear fashion, $\mathbf{\mathcal{G}}$ a Laplacian connecting
matrix, $\mathbf{\mathcal{C}}$ an adjacent connecting matrix,
$\mathbf{\sigma}$ the strength of the linear coupling and $g$ the
strength of the non-linear coupling. Finally, $N$ is the number of
nodes.

We say that a network is of the class UPPER if $\Lambda \geq
\Lambda_C$ and of the class LOWER if $\Lambda \leq \Lambda_C$.

We consider that the UPPER and LOWER property holds for a properly
rescaled coupling strength intervals
$\sigma(N,\mathcal{G},\mathcal{C})$ $\in$
$[\sigma_m(N,\mathcal{G},\mathcal{C}),\sigma^{*}(N,\mathcal{G},\mathcal{C})]$
and $g(N,\mathcal{G},\mathcal{C})$ $\in$ ${\ \ }$
$[g_m(N,\mathcal{G},\mathcal{C}),g^{*}(N,\mathcal{G},\mathcal{C})]$.

{\bf \it Conjecture}: {\it The LOWER or UPPER character of a network
  described by Eq. (\ref{element_dynamics}) is independent of the
  number of nodes for a properly rescaled coupling strength interval.}

In simple words, this conjecture states that as long as one preserves
the coupling functions $\mathbf{H,S}$ under which nodes connect among
themselves, there will be coupling strengths $\sigma,g$ for which the
LOWER or UPPER character of an active network will be preserved,
regardless of the number of nodes $N$.

\section{Defining the coupling strength intervals}

For simplicity in the notation, we ommit in the representation of the
constants $\sigma_m,\sigma^{*}$ and $g_m,g^*$ the reference to their
dependence on $\mathbf{\mathcal{G},\mathcal{C}}$.

Our conjecture only states that whenever there is a network with $N_1$
nodes with a structure defined by
$\mathbf{H,S,\mathcal{G},\mathcal{C}}$ and this network has an UPPER
(or lower) character for the coupling strength intervals
$[\sigma_m(N_1),\sigma^{*}(N_1)]$ and $[g_m(N_1),g^*(N_1)]$ then if a
network with $N_2$ nodes is constructed preserving the coupling
functions $\mathbf{H,S}$ then there exists coupling strength intervals
$[\sigma_m(N_1),\sigma^{*}(N_1)]$ and $[g_m(N_1),g^*(N_1)]$ for which
the network behaves with the same UPPER (or lower) character.
 
To make this conjecture more practical, we make in the following some
assumptions.

The value of the constants $\sigma_m(N),\sigma^{*}(N)$ and
$g_m(N),g^*(N)$ are such that either ${\ \ }$ $|\sigma_m(N)/g_m(N)|$ $>>$ 1 or
$|\sigma_m(N)/g_m(N)|$ $<<$ 1 and $|\sigma^*(N)/g^*(N)|$ $>>$ 1 or
$|\sigma^*(N)/g^*(N)|$ $<<$ 1. The reason is because for such conditions,
the values for these constants for a network with $N>2$ nodes can be
calculated from the values of these constants from the reference
network, in here assumed to have $N=2$ nodes.

The network with $N_1$ nodes is regarded to be the {\it reference}
network and we consider that $N_2>N_1$. For simplicity, we further
consider that $N_1=2$. In addition, to make our analyses simpler, we
consider in our numerical simulations a constant $g_m(N)=g^*(N)$, and
we choose either $|\sigma_m(N)/g_m(N)|>>$1 or
$|\sigma_m(N)/g_m(N)|<<$1.

Then, we choose the constant $\sigma^{*}(N)$ such that
its value is a little bigger than the smallest coupling values for
which complete synchronization is reached and when $\Lambda =
\Lambda_C$. However, other intervals could be considered. The reason again is that
$\sigma^{*}(N)$ can be analytical calculated from $\sigma^{*}(N=2)$,
the linear coupling strength, for which complete
synchronization is found in two mutually coupled systems.

The constants that define the coupling strength interval for a network
with $N$ nodes can be calculated from the constants that define the
coupling strength interval for a network with $N=2$ nodes using

\begin{eqnarray} 
\sigma(N)=\frac{2\sigma(N=2)}{|\gamma_2(N)|} \label{rescale1} \\ 
g(N)=\frac{g(N=2)}{k} \label{rescale2}
\end{eqnarray}
\noindent
where $\gamma_{2} $ is the second largest eigenvalue of $
\mathcal{G}$, and $k$ is the number of incoming connections of each
node of the network.

As an example of how we use Eq. (\ref{rescale1}), we do the
following. Having defined that two mutually linearly coupled systems
(so, $g$=0) have a LOWER character for the linear coupling strength
interval $[\sigma_m(N=2),\sigma^*(N=2)]$,then we construct a network
using the same linear coupling function composed of $N$ nodes, but
considering now the linear coupling strength interval
$[\sigma_m(N),\sigma^*(N)]$ calculated using Eq. (\ref{rescale1}).
According to our conjecture, such a network will have a lower
character.

For a more detailed analysis of how we derive Eqs. (\ref{rescale1})
and (\ref{rescale2}), one should read Appendix \ref{couples}.

\section{Networks of coupled maps}

Here, we consider only linear couplings. Then $g=g_m$=0, and
therefore, $\sigma_m=0$.

For general networks (discrete or continuous descriptions) whose nodes
are completely synchronous, one always have that $\Lambda=\Lambda_C$,
a non generic case for which our conjecture can be proved.

For networks of coupled maps, there is another trivial example when
$\Lambda=\Lambda_C$. That happens for networks whose Jacobian is
constant as networks formed by linear maps of the type $x^{(i)}_{n+1}
= \alpha x^{(i)}_{n} + 2 \sigma \sum_{j=1}^N
\mathcal{G}_{ij} x^{(j)}_n $ (mod 1) and when there exists complete
synchronization, and the attractor lays on the synchronization
manifold. These results concern arbitrary connecting Laplacian
matrices $\mathcal{G}_{ij}$, for example, they would apply for map
lattice with a coupling whose strength decreases with the distance as
a power-law \cite{viana}.

\begin{figure}[!h]
\centerline{\hbox{\psfig{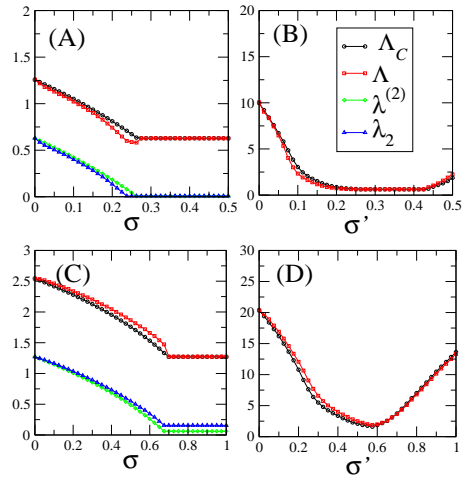}}} \caption{Results for the network in
   Eq. (\ref{mapa1}), for $\rho$=0.5. For (A) and (C), N=2, and for
   (B) and (D), $N$=16. An inhibitory (UPPER) network is shown in (A) and
   (B), for $s$=-1, and an excitable (LOWER) network is shown in (C) and
   (D), for $s$=1. The horizontal axis in (B) and (D) were rescaled by
   $\sigma^{\prime}$=$\sigma*|\gamma_2(N=16)|/2$, so that one can
   compare Figs. (B) and (D) with (A) and
   (C).$|\gamma_2(N=16)|$=4.1542.}
\label{mapa_01}
\end{figure}          

Now, imagine the following network {\small
\begin{equation}
x_{n+1}^{(i)}=2x_n^{(i)}+ s \rho {x_n^{(i)}}^2 + 2 \sigma
\sum_{j=1}^N \mathcal{G}_{ij} x^{(j)}_n \text{ (mod 1)}
\label{mapa1}
\end{equation}}
\noindent
with $\rho \geq 0$ and $s=\pm 1$.  The synchronization manifold
is defined by $x_n^{(1)}=x_n^{(2)}=\ldots=x_n^{(N)}$, and in an
all-to-all connecting topology, the Lyapunov exponent of the
synchronization manifold can be calculated by $\lambda^{(1)}=\ln{(2)}
+ 1/t \sum_n
\ln{|1 + s \rho x_n|}$, with $n=(1,\ldots,t)$, and the others $N-1$ 
equal exponents associated to
the transversal directions by $\lambda^{(i)}=\ln{(2)}+1/t\sum_n
\ln{|1+s \rho x_n -2\sigma|}$, for $i\geq 2$. In Fig.
\ref{mapa_01}, we show the values of $\Lambda$ and $\Lambda_C$
as we vary $\sigma$, for $\rho=0.5$. In (A) and (C), we consider $N$=2
(all-to-all topology), and in (B) and (D) we consider a random
networks formed by $N$=16 nodes. The coupling strength interval used
for two coupled nodes was rescaled to the proper coupling strength
interval for the larger random network, using in the denominator of
Eq. (\ref{rescale1}) the value of $|\gamma_2|=4.1542$, relative to the
second largest eigenvalue (in absolute value) of the random
network. One can check that if two coupled nodes have an UPPER [LOWER]
character for a given coupling interval as can be seen in
Fig. \ref{mapa_01}(A) [in Fig. \ref{mapa_01}(C)], larger networks will
behave in the same UPPER [LOWER] character as can be seen in Fig.
\ref{mapa_01}(B) [in Fig. \ref{mapa_01}(D)]. 

The conjecture describes a relationship between the conditional
exponents and the Lyapunov exponents. To see that, notice that,
typically for the UPPER networks of linearly connected maps, we have
$\lambda_1 \approx \lambda^{(1)}$, a consequence of the fact that the
largest Lyapunov exponent can be calculated using the same directions
as the ones along the synchronization manifold.  Thus, using our
conjecture, if the network is of the UPPER type, $\lambda_1+\lambda_2
\leq
\lambda^{(1)}+\lambda^{(2)}$, which provides $\lambda_2 \leq
\lambda^{(2)}$. Otherwise, if the network is of the LOWER type,
$\lambda_2 \geq \lambda^{(2)}$. That can be checked in
Figs. \ref{mapa_01}(A)-(C). Since the approaching of the transversal
conditional exponents to negative values are associated with the
stabilization of a certain oscillation mode, close to a coupling
strength for which a transversal conditional exponent approaches zero,
there will also be a Lyapunov exponent which approaches zero, meaning
that some oscillation in the attractor becomes stable.

\section{Networks of Hindmarsh-Rose neurons}

Let us illustrate our conjecture in networks composed of $N$ coupled
Hindmarsh-Rose neurons \cite{HR} electrically and chemically coupled
\cite{comment_hussein}: {\small
\begin{eqnarray}
\dot{x}_i &=& y_i + 3x_i^2 - x_i^3-z_i + I_i +
g\sum_{j=1}^{N}\mathcal{C}_{ij}S(x_i, x_j) \nonumber \\ && + \sigma
\sum_{j=1}^{N}\mathcal{G}_{ij}x_j \label{HR}\\
\dot{y}_i &=& 1-5x_i^2-y_i; \:\:\: \dot{z}_i  = -rz_i +
4r(x_i+1.6),  \nonumber
\end{eqnarray}
} \noindent The parameter $r$ modulates the slow dynamics and is
set equal to 0.005, such that each neuron is chaotic.
The synaptic chemical coupling is modeled by $S(x_i,
x_j)=(x_i-V_{syn})\Gamma(x_j)$ where $
\Gamma(x_j)=\displaystyle\frac{1}{1+e^{-\theta(x_j
-\Theta_{syn})}}$ with $\Theta_{syn}=-0.25$, $\theta=10$ and
$V_{syn}=2.0$.  $\sigma {\bf \mathcal{G}}_{ji}$ is the strength of the
electrical coupling between the neurons, and $I_i=3.25$. In order to
simulate the neuron network and to calculate the Lyapunov exponents
through Eq. (\ref{variational}), we use for the node $i$ the initial
conditions $x_i$=-1.3078+$\omega_i$, $y_i$=-7.3218+$\omega_i$, and
$z_i$=3.3530+$\omega_i$, where $\omega_i$ is an uniform random number
within [0,0.02]. To calculate the conditional exponents
$\lambda^{(i)}$, we use in Eq. (\ref{variational1}) the initial
conditions, $x$=-1.3078, $y$=-7.3218, and $z$=3.3530, but any other
set of typical equal initial conditions can be used
\cite{comenta_conjectura}.

\begin{figure}[!h]
 \centerline{\hbox{\psfig{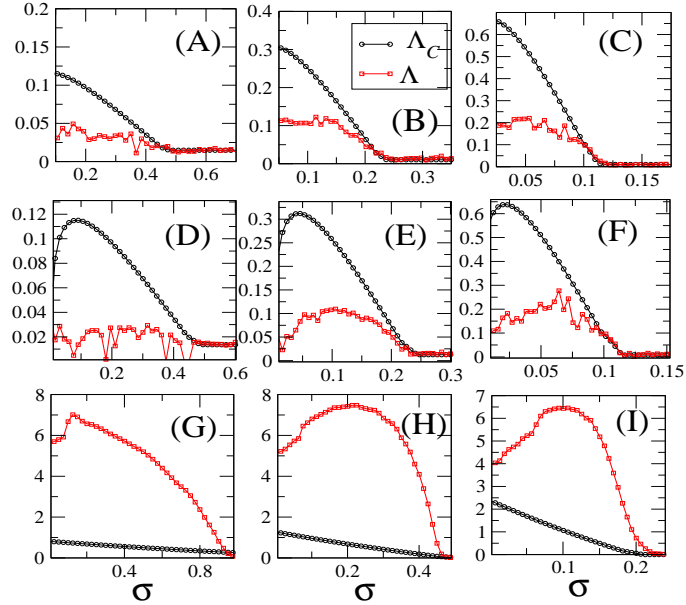}}} 
\caption{The values of $\Lambda$ and
 $\Lambda_C$ for neural networks described by Eq. (\ref{HR}) of nodes
 connected in an all-to-all topology.  In (A),(D), and (G), $N$=2. In
 (B),(E), and (H), $N=4$. In (C), (F), (I), $N$=8. Results for
 networks with an UPPER character are shown in (A-F), and for networks
 with an LOWER character are shown in (G-I).}  \label{neuron}
 \end{figure}

We study three types of neural networks. (i) $g<0$ [Figs.
\ref{neuron}(A-C)]. The coupling (synapses) is said to be of the
excitatory type, since $x_i-V_{syn} < 0$ and the nodes $j$ contribute
positively in the equations for the first derivative of $x_i$. In
other words, the postsynaptic neuron ($x_i$) is forced to opposite the
presynaptic ones ($x_j$); (ii) $g=0$ [Figs.  \ref{neuron}(D-F)]. The
network has nodes coupled to other nodes only electrically.  From the
biological point of view, neurons only make electrical connections
with their nearest neighbors.  Here, we also consider long-range
correlations. Since $\sigma \geq 0$, this coupling contributes
negatively to the first derivative of $x_i$, which results in an
inhibitory effect to the oscillatory motion of the neuron $x_i$.
(iii) $g>0$ [Figs.  \ref{neuron}(G-I)]. The coupling (synapses) is
said to be of the inhibitory type, since the nodes $j$ contribute
negatively in the equations for the first derivative of $x_i$. For
such a case, the postsynaptic neuron ($x_i$) is forced to synchronize
its rithmus to the rithmus of the presynaptic ones ($x_j$).

In Fig. \ref{neuron}, we show the values of $\Lambda$ and $\Lambda_C$
for the three types of neural networks being considered, case (i) in
Figs. \ref{neuron}(A-C), case (ii) in Figs. \ref{neuron}(D-F), and
case (iii) in Figs. \ref{neuron}(G-I). Networks whose results are
represented in Figs.  \ref{neuron}(A-C) and (G-I) are constructed by
neurons connected simultaneously electrically ($\sigma > 0$) and
chemically ($g > 0$) in the all-to-all topology, while networks whose
results are represented in Figs.  \ref{neuron}(D-F) are constructed by
neurons connected only electrically ($\sigma > 0$ and $g$=0) in the
all-to-all topology.

In (A) [case (i)], for $N$=2 and $g=-0.01$, $\Lambda\leq \Lambda_C$,
for $\sigma=[0.1,0.7]$. So, $\sigma_m(N=2)$=0.1 which leads to
$\sigma_m(N=2)/g_m(N=2)|>>1$, as we wish. From our conjecture, for
larger networks as the ones shown in Figs. \ref{neuron}(B) [$N=4$] and
\ref{neuron}(C) [$N$=8], we must have $\Lambda \leq \Lambda_C$, for
the rescaled coupling interval. From Eqs. (\ref{rescale1}) and
(\ref{rescale2}), we have for the network with $N=4$
[Fig. \ref{neuron}(B)], the rescaled coupling strength interval should
be $\sigma=[0.1/2,0.7/2]$ and $g=-0.01/3$, and for the network with
$N=8$ [Fig. \ref{neuron}(C)], the rescaled coupling strength interval
should be $\sigma=[0.1/4,0.7/4]$ and $g=-0.01/7$. In fact, as one sees
in Figs.
\ref{neuron}(B-C), we indeed see that these networks have the same
UPPER character as the network with $N$=2, for the considered coupling
strength intervals.

In (D) [case (ii)], for $N$=2 and $g=0$, $\Lambda\leq \Lambda_C$ for
$\sigma=[0,0.6]$. So, $g_m(N=2)=0$ and consequently $\sigma_m(N=2)=0$.
From our conjecture, for larger networks as the ones shown in
Figs. \ref{neuron}(E) [$N=4$] and \ref{neuron}(F) [$N$=8], we must
have $\Lambda\leq \Lambda_{C}$ for the rescaled coupling
interval. From Eqs. (\ref{rescale1}) and (\ref{rescale2}), we have for
$N=4$ [Fig. \ref{neuron}(E)], the rescaled coupling interval should be
$\sigma=[0,0.6/2]$ and for $N=8$ [Fig. \ref{neuron}(F)], the rescaled
coupling interval should be $\sigma=[0,0.6/4]$.  In fact, as one sees
in Figs.  \ref{neuron}(E-F), we indeed have that these networks have
the same UPPER character of the network with $N$=2.

Finally, In (G) [case (iii)], for $N$=2 and $g$=10, $\Lambda \geq
\Lambda_C$ for $\sigma=[0.01,1]$. So, $|g_m(N=2)/\sigma_m(N=2)|>>1$ as we
wish. From our conjecture, for larger networks, as the ones shown in
Figs. \ref{neuron}(H) [$N=4$] and
\ref{neuron}(I) [$N$=8], we must have $\Lambda\geq \Lambda_{C}$ for
the rescaled coupling interval. From Eqs. (\ref{rescale1}) and
(\ref{rescale2}), and $N=4$ [Fig. \ref{neuron}(H)], the rescaled
coupling interval should be $\sigma=[0.01/2,1/2]$ and $g$=10/3, and
for $N=8$ [Fig. \ref{neuron}(I)], the rescaled coupling interval
should be $\sigma=[0.01/4,1/4]$ and $g$=10/7.  In fact, as one see in
Figs.
\ref{neuron}(G-I), we indeed have that these networks have the same
LOWER character of the network with $N$=2.

An inhibitory chemical coupling inhibits the nodes of the network,
which means that such a coupling forders an increase in the level of
synchronization.  On the other hand, an excitatory chemical coupling
excites the nodes, which means that such a coupling forders an
increase in the level of desynchrony. 

It is intuitive to imagine that an excitatory network (as defined
exclusively in terms of the chemical coupling) would have a LOWER
characteristic and an inhibitory network (as defined exclusively in
terms of the chemical coupling) would have an UPPER characteristic.
That is why excitation would mean an increase of desorganization (more
entropy) and inhibition an increase of synchronization (less entropy).
However, we have previously shown in Figs. \ref{neuron}(A-C) that an
excitatory network (as usually defined in terms of the chemical
coupling) has the UPPER characteristic and in Figs. \ref{neuron}(G-I)
that an inhibitory network (as usually defined in terms of the
chemical coupling) has the LOWER characteristic. This aparent
contradiction is simple to be explained.

In the excitatory networks [Figs.  \ref{neuron}(A-C)], the absolute
strength of the non-linear (chemical) coupling (0.01) is smaller than
the strength of the linear (electrical) coupling. As a consequence,
the linear coupling prevails on the non-linear coupling.  In the
inhibitory networks [Figs.  \ref{neuron}(A-C)], the strength of the
non-linear (chemical) coupling (10) is much larger than the strength
of the linear coupling.  However, such a large strength effectively
forders an excitatory behavior in the network.  Notice that while in
Fig.
\ref{neuron}(A) complete synchronization appears for $\sigma \approx
0.5$, in Fig.  \ref{neuron}(G) complete synchronization appears for
$\sigma \approx 0.95$, and therefore, complete synchronization in the
inhibitory network appears only for a larger linear coupling than the
one for which complete synchronization appears in the excitatory
networks.

It is not the scope of this work to determine for which conditions an
inhibitory (or excitatory) non-linear (chemical) couplings in networks
of neurons simultaneously connected by linear and non-linear means
determines the UPPER or LOWER character of a network. For that one
should check Ref.  \cite{francois}. Had we consider that the neurons
were connected exclusively by non-linear (chemical) means
($\sigma=0$), then it is to be expected that inhibitory networks would
present an UPPER character and excitatory networks would present a
LOWER character.

\section{Application of our conjecture to predict the chaotic behavior of large
networks}

\begin{figure}[!h]
  \centerline{\hbox{\psfig{file=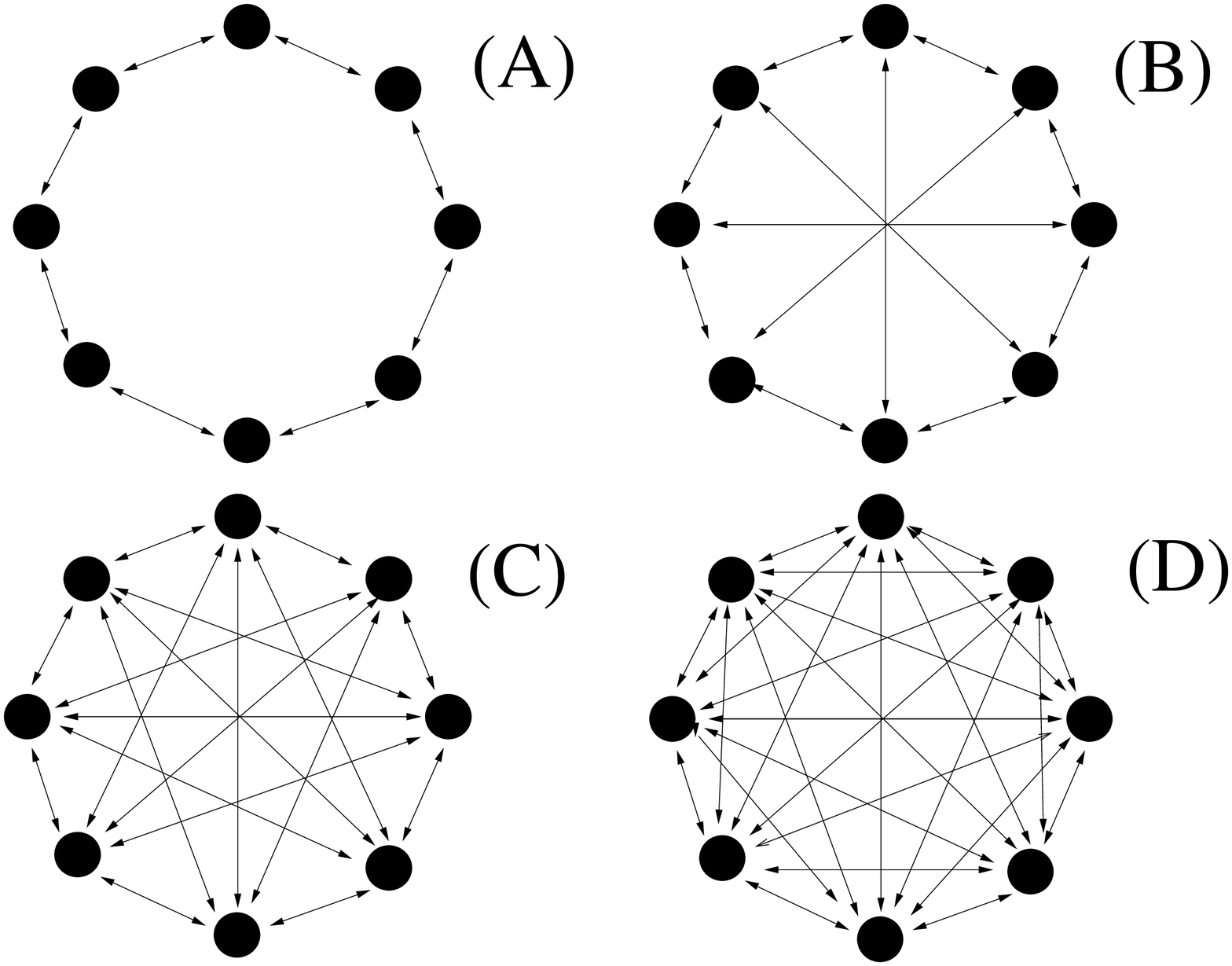,height=5cm,width=6cm}}} 
\caption{Representation of a few network
  topologies with 8 neurons, considered in this work. The filled balls
  represent neurons and the lines indicate an electric bidirectional
  coupling. In (A) the neurons are only coupled with its nearest
  neighbors, forming a ring.  From (B) to (D) it is added to the
  network long-range bidirectional connections, The average number of
  connections that each neuron receives (network degree), $\omega$, is
  $\omega$=2, in (A), $\omega$=3, in (B), $\omega$=5, in (C), and
  $\omega$=7, in (D).  In a network with $N$ neurons, long-range
  connections are introduced in the initial ring by connecting each
  neuron to its N/2-th (B) neighbors, then to its ($N/2$-1)-th
  neighbors (C), then to its ($N/2-l$)-th neighbors, till each neuron
  is connected to its second neighbors, when the network has the
  all-to-all coupling topology.}
 \label{paper_HR_fig2}
 \end{figure}

In the following, we discuss how our conjecture can be used to make
general statements about active networks. Consider the UPPER networks
formed by neurons connected only electrically ($g$=0). For such cases,
$\Lambda_C(N)$ is an upper bound for the Kolmogorov-Sinai (KS) entropy
$H_{KS}$ (see
\cite{comentario}) and also an upper bound for $\Lambda$. Since networks
formed by nodes connected in an all-to-all topology produce Laplacian
matrices whose eigenvalues are $\gamma_1=0$, and $\gamma_i$=$-N$, for
$i=2,\ldots,N$, it is clear from Eq. (\ref{rescale1}) that
$\max{[\Lambda_C(N)]}$ for the considered coupling strengths of a network
with the all-to-all topology, is larger or equal to $\max{[\Lambda_C(N)]}$
for any other topology. Defining the {\it network capacity},
$c(N)$, to be equal to $\max{[\Lambda_C(N)]}$, calculated for
the all-to-all topology (and the considered coupling intervals), since
$\Lambda_C(N) \geq \Lambda(N)$ (as well as $\Lambda_C(L) \geq H_{KS}(N)$
\cite{ruelle}) for UPPER networks, we conclude that for these networks
not only {\small
\begin{equation}
c(N) \geq \max{[\Lambda(N)]}
\label{capacidade}
\end{equation}
}
\noindent
but also 
{\small
\begin{equation}
c(N) \geq \max{[H_{KS}(N)]}
\end{equation}
}
\noindent
where the $\max$ of $\Lambda(N)$ in taken considering "any" possible
topologies (described in Fig. \ref{paper_HR_fig2}) and the considered
coupling intervals.

The value of $c(N)$ for neural networks electrically
connected can be approximately calculated by $\max{(\lambda^{(1)})} +
(N-1)\max{(\lambda^2)}$ (notice that since ${\lambda^{(1)}}$ does not
depend on $\sigma$, then, $\max{(\lambda^{(1)})}$ happens for the same
coupling strength for which $\max{(\lambda^{(2)})}$ is found), which
leads to $c(N) \cong 0.01362 + 0.1013(N-1)$ bits/(time
unit).  By doing simulations considering networks as the ones
represented in Fig. \ref{paper_HR_fig2}, (with $10 \leq N \leq 40$),
we obtain that $\max{[\Lambda(N)]} \cong 0.0830 +
0.0230(N-1)$bits/(time unit), which agrees with
Eq. (\ref{capacidade}).

For a network with the all-to-all topology [as in
Fig. \ref{paper_HR_fig2}(D)], for $N\geq10$, we obtain $\max{[\Lambda(N)]}
\cong 0.158447 + 0.031537(N-1)$, which agrees with Eq. (\ref{capacidade}), 
because $c(N) \geq
\max{[\Lambda(N)]}$ (where the maximum is taken considering the all-to-all
topology). Finally, if we construct a network with nodes connecting to their
nearest neighbors forming a closed ring [as in Fig. \ref{paper_HR_fig2}(A)],
we find $\max{[\Lambda(N)]} \cong 0.197125 + 0.034865(N-1)$bits/(time
unit). Equation (\ref{capacidade}) is once again verified.

Thus, $c(N)$ for electrically connected networks does not depend
on the network topology. That is not the case for chemically connected neural
networks, for which $c(N)$ might be achieved for different
topologies, since the curve for $\lambda^{(1)}$ and $\lambda^{(i)}$ achieve
their maximal values for different values of the coupling strength.

Further, consider two coupled LOWER-type systems and $\Lambda$ is null (positive)
for some coupling strength, meaning a periodic behavior (meaning chaos). It
might be that, for a proper rescaled coupling strength, as more nodes are
added to the network, $\Lambda$ becomes positive, meaning chaos (for sure there
will be chaos). We can also use our conjecture to predict the behavior of a
network constructed with nodes that are either chaotic or periodic, by only
having information about two coupled nodes. Considering only linear couplings
[$g$=0, in Eq. (\ref{element_dynamics})]. For $\sigma \leq
\epsilon$, the two coupled nodes have a periodic dynamics, and thus,
$\Lambda=0$, but $\Lambda_C >0$ (UPPER character). That implies that as we
add more nodes in the network, it might be that after the proper rescaling of
the coupling strength the network becomes chaotic.

\section{Conclusions}

In conclusion, we have presented arguments to suggest that for a class
of dynamical systems, the sum of all the positive Lyapunov exponents
of an active network is bounded by the sum of all the positive
Lyapunov exponents of the synchronization manifold.  In practical
terms, the entropy production of the synchronization manifold and its
transversal directions ($\Lambda_C$) of a system of two coupled
equal dynamical systems determines the upper (LOWER character) or lower
(UPPER character) bound for the sum of the positive Lyapunov exponents of
a large network. This fact enables one to predict the behavior of a
large network by using information provided by only two coupled nodes.

Our results indicate that the behavior (synchronization and
information) of an active network with nodes possessing equal dynamics
and especial properties \cite{comenta_conjectura} does not strongly
depend on the coupling topology (${\mathcal{G}}$ and ${\mathcal{C}}$)
and the size of the network ($N$) but rather on the nature of the
coupling functions ($S$ and $H$).
 
At first glance, this result seems to be in direct conflict with what
one would expect to find in realistic neural networks, as the
mammalian brain, whose topology is possibly responsible for
intelligence. But one should have in mind that the here considered
networks are constructed with nodes that possess equal dynamics being
connected using always the same coupling function. In realistic brain
networks, the coupling functions largely differ along different brain
areas as well as the coupling strength depends on time. Therefore, in
order for the topology to play an important role in the behavior of a
network one needs to consider networks with non-equal nodes and/or
that possess coupling functions that change in space and time.

Naturally, the large class of networks for which our conjecture
applies are far from being realistic. However, we believe our
conjecture can contribute to the understanding of much more complex
networks.  For example, for the UPPER networks, a large series of
numerical results show that more realistic networks constructed with
non-equal nodes (or networks of equal nodes but with random coupling
strengths \cite{PLOSONE}) have a KS entropy smaller than the networks
with equal nodes. Therefore, even though networks with equal nodes
might not be realistic, their entropy production is an upper bound for
the entropy production of more realistic networks.

Excitability and inhibition is a concept usually used to classify the
way non-linear (chemical) synapses between two neurons are done. When
an inhibitory neuron spikes (the pre-synaptical neuron) a neuron
connected to it (the post-synaptical neuron) is prevented to spike.
When an excitatory neuron spikes it induces the post-synaptical neuron
to spike. Intuitively, one should expect that an inhibitory
(excitatory) coupling forders (prevents) synchronization, but we have
shown two cases for which an excitatory network had an UPPER character
and an inhibitory network had a LOWER character. The reason is that
only the non-linear couplings (chemical synapses) are not sufficienty
to define the LOWER and UPPER character of the network. One should
consider the combined effect of the linear (electrical) and of the
non-linear couplings (chemical). For more details about that, see Ref.
\cite{francois}

For UPPER networks, the entropy of the attractors cannot be larger
than the entropy of the synchronous set, which therefore imposes a
clear limit in the complex character of these networks. On the hand,
for LOWER networks, our conjecture states that such a limit is
unknown.

This conjecture might be a consequence of the fact that the attractors
and behaviors that appear in two coupled nodes for a given coupling
strength are similar to the ones that appear for larger networks, to
parameters rescaled according to Eqs. (\ref{rescale1}) and
(\ref{rescale2}). In fact, as one can see in the work
\cite{baptista_PRE2008}, that is indeed the case for the coupling
strengths for which burst phase synchronization (BPS) or phase
synchronization (PS) appear in networks of electrically coupled
HR-neurons.

\section{Appendix}

\subsection{The conjecture}\label{conjecture}

Let $ \mathbf{H,S,\mathcal{G},\mathcal{C}},\sigma,g,N$ as in
(\ref{element_dynamics}) to be the parameters which defines the active
network.  $\mathbf{H}$ represents the function under which the nodes
connects among themselves in a linear fashion, $\mathbf{S}$ the
function under which the nodes connects among themselves in a
non-linear fashion, $\mathbf{\mathcal{G}}$ a Laplacian connecting
matrix, $\mathbf{\mathcal{C}}$ an adjacent connecting matrix,
$\mathbf{\sigma}$ the strength of the linear coupling and $\mathbf{g}$
the strength of the non-linear coupling. Finally, $N$ is the number of
nodes.

Denote by $\Lambda(\mathbf{H,S,\mathcal{G},\mathcal{C}},\sigma,g,N)$
and $ \Lambda_C(\mathbf{H,S,\mathcal{G},\mathcal{C}},\sigma,g,N)$ the
sum of the posititve Lyapunov exponents and the sum of the positive
conditional Lyapunov exponents of the network whose structure is
specified by $ (\mathbf{H,S,\mathcal{G},\mathcal{C}}) $,
respectively. We say that the couple $(\mathbf{H,S})$ makes the
network to be of the LOWER class if for every $
(\mathbf{\mathcal{G},\mathcal{C}}) $ there exist four positive
constants $ \sigma_m$, $g_m$, $\sigma^{*}$ and $g^{*}$ such that
\begin{equation}
	\label{conjecture1}
	\Lambda_C(\mathbf{H,S,\mathcal{G},\mathcal{C}},\sigma,g)
	\geq \Lambda(\mathbf{H,S,\mathcal{G},\mathcal{C}},\sigma,g)
\end{equation} 
for all $ \sigma_m \le \sigma \le \sigma^{*} $ and all $ g_m \le g \le
g^{*} $. An UPPER class active network is defined similarly by
reversing the direction of inequality (\ref{conjecture1}).

{\bf \it Conjecture: } Given a network with a LOWER (UPPER) character
[as defined in (\ref{conjecture1})] specified by
$(\mathbf{H,S,\mathcal{G},\mathcal{C}})$, and
$(\mathbf{\mathcal{G},\mathcal{C}}) $ with $N_1$ nodes, there exist
coupling strength intervals $ \tilde{\sigma_m} \le \sigma \le
\tilde{\sigma^{*}} $ and $ \tilde{g_m} \le g \le
\tilde{g^{*}}$ for which a network specified by 
$(\mathbf{H,S,\mathcal{\tilde{G}},\mathcal{\tilde{C}}})$ and
$(\mathcal{\tilde{G}},\mathcal{\tilde{C}})$ with $N_2$ nodes has also
a LOWER (UPPER) character.

\subsection{Derivation of the coupling strength constants}\label{couples}

The variational equation (\ref{variational0}) for
the synchronous solution can be written as follows
\begin{eqnarray}
	\label{variational}
\delta\mathbf{\dot{X}} & = & \{\mathbf{I}\otimes D\mathbf{F(\mathbf{x})}+\sigma {\mathcal{G}} \otimes D \mathbf{H(\mathbf{x})} + g {\mathcal{C}} \otimes D_{1} \mathbf{S(\mathbf{x},\mathbf{x})} \nonumber \\
& + & g k \mathcal{C}\otimes D_{2}
\mathbf{S(\mathbf{x},\mathbf{x})}\}\delta\mathbf{X},
\end{eqnarray}
where $ \mathbf{\delta X} $ is the column vector of $ \R^{Nd} $ with components  $ \mathbf{\delta x}_1,\mathbf{\delta x}_2,\ldots,\mathbf{\delta x}_N $, and $\otimes$ stands for the Kronecker product of matrices. 
Since $\mathcal{G}$ and ${\mathcal{C}}$ commute, they can be simultaneously diagonalized. Let $ \mathbf{u}_{1},\ldots,\mathbf{u}_{N} $ be their eigenvectors, and denote by $ \gamma_{1},\ldots,\gamma_{N} $ and $ \tilde{\gamma}_{1},\ldots,\tilde{\gamma}_{N} $ the corresponding eigenvalues for $ \mathcal{G} $ and $ \mathcal{C} $, respectively. We order $ \{\gamma_{i}\} $ so that $ \gamma_{1} = 0 $. If we write $ \mathbf{\delta X}(t)=\sum_{1 \le i \le N} \mathbf{u}_{i} \otimes \mathbf{y}_{i}(t) $ with $ \mathbf{y}_{i}(t) \in \R^{d} $, and substitute it in (\ref{variational}), then a straightforward computation gives
\begin{equation}
	\label{variational1}
\dot{\mathbf{y}_i} = \{D \mathbf{F(\mathbf{x})}+\sigma \gamma_i D \mathbf{H}(\mathbf{x}) + g k D_{1} \mathbf{S(\mathbf{x},\mathbf{x})}+ g \tilde{\gamma}_i D_{2}\mathbf{S(\mathbf{x},\mathbf{x})}\} \mathbf{y}_i.
\end{equation}

While Eq. (\ref{variational}) describes how perturbations are
propagated or damped along a particular node of the network
($\bf{x}_i$) Eq. (\ref{variational1}) describes how perturbations are
propagated along an eigenmode ($\mathbf{y}_i$). While Eq.
(\ref{variational}) is valid for networks with nodes initially set in
typical initial conditions Eq. (\ref{variational1}) is only valid for
networks with nodes initially set with equal initial conditions, the
assumption done in order to place Eq. (\ref{variational}) in the
eigenmode form in Eq. (\ref{variational1}).

Calculating the Lyapunov exponents from Eq. (\ref{variational0})
assuming equal initial conditions for every node provides the same
exponents than the conditional ones obtained from Eq.
(\ref{variational1}). An advantage of using Eq. (\ref{variational1})
for the calculation of the conditional exponents is that while Eq.
(\ref{variational}) requires the employement of $Nd \times Nd$
dimensional matrices, the conditional exponents by Eq.
(\ref{variational1}) requires the use of $N$ matrices of
dimensionality $d$. A mode $i$ in equation in Eq. (\ref{variational1})
provides a set of $d$ conditional exponents, denoted by
$\lambda^{(i)}_j$, $j=1,\ldots,d$. Since we are only interested in
positive exponents, we simplify the notation by making $\lambda^{(i)}
= \sum_{j=1}^d \lambda^{(i)}_j$. So, $\lambda^{(1)}$ refers to the sum
of the positive conditional Lyapunov exponents of the synchronization
manifold while $\lambda^{(i)}$ ($i\ge 2$) refer to the sum of the
positive Lyapunov exponents of the transversal directions to the
synchronization manifold.

From Eq.  (\ref{variational1}) it becomes clear that once the
conditional exponents are calculated using two bidirectionally coupled
nodes, for the considered coupling interval, the conditional exponents
of the mode $i$ ($\lambda^{(i)}$) for larger networks with arbitrary
topology can be calculated from the exponents for $N$=2, by
$\lambda^{(1)}(N=2,\sigma,g)=\lambda^{(1)}(N,\sigma,g/k)$ and
$\lambda^{(2)}(N=2,\sigma,g)=\lambda^{(i)}(N,2\sigma/|\gamma_i(N)|,g/k)$.

To understand why, just make in Eq. (\ref{variational1}) $g=0$. The
only term that changes in these equations as one considers networks
with different topologies and sizes is $\gamma_i(N)$, the $i-th$
eigenvalue of the connecting Laplacian matrix ${\mathcal{G}}$ with
size $N$. Denoting $\gamma_i(N=2)$ and $\sigma(N=2)$ to be the $i-th$
eigenvalue of the Laplacian matrix ${\mathcal{G}}$ and the coupling
strength, respectivelly, for two mutually coupled nodes then the mode
$i$ of Eqs. (\ref{variational1}) for a network with a number $N$ of
nodes will preserve the form of the mode $i$ in Eqs.
(\ref{variational1}) for the network with $N=2$ if
$\sigma(N)=2\sigma(N=2)/|\gamma_i(N)|$. For practical purposes, this
relation can be expressed in terms of only the coupling strengths.
Denoting $\tilde{\sigma}$ as the strength value for the linear coupling
for which $\lambda^{(2)}(N=2)$ reaches a given value,
then the coupling strengths for which $\lambda^{(i)}(N)$ reaches the
same value is given by the rescaling \cite{francois} {\small

$$
\tilde{\sigma}(N)=\frac{2\tilde{\sigma}(N=2)}{|\gamma_i(N)|}
$$

A similar analysis can be done assuming that $\sigma$=0. Once that
$D_{2}\mathbf{S(\mathbf{x},\mathbf{x})} <<
D_{1}\mathbf{S(\mathbf{x},\mathbf{x})}$ in Eq. (\ref{variational1}),
then the only term that changes in these equations as one considers
networks with different topologies and sizes is $k(N)$, the number of
connections a node within a network of $N$ nodes receives from the
other nodes. So, denoting $\tilde{g}$ as the strength values for the
non-linear coupling for which $\lambda^{(2)}(N=2)$
reaches a given value, then the coupling strength for which
$\lambda^{(i)}(N)$ reaches the same value is given by the rescaling
\cite{francois} {\small

$$
\tilde{g}(N)=\frac{\tilde{g}(N=2)}{k}
$$

As shown in Ref. \cite{francois}, Eqs. (\ref{rescale1}) and
(\ref{rescale2}) remain valid if either
$|\tilde{\sigma}/\tilde{g}|>>1$ or $|\tilde{g}/\tilde{\sigma}|>>1$,
which means that one can consider the linear coupling as a
perturbation ($|\tilde{g}/\tilde{\sigma}|>>1$) or the nonlinear
coupling as a perturbation ($|\tilde{\sigma}/\tilde{g}|>>1$).

Further in this work, the coupling interval is rescaled using as a
reference the second largest conditional exponent $\lambda^{(2)}$
computed for the network with $N$=2.

\textbf{Acknowledgment} MSB acknowledges the partial financial support
of "Funda\c c\~ao para a Ci\^encia e Tecnologia (FCT), Portugal"
through the programmes POCTI and POSI, with Portuguese and European
Community structural funds. This work is also supported in part by the
CNPq and FAPESP (MSH). MSH is the Martin Gutzwiller Fellow 2007/2008.

\begin{thebibliography}{99}
\bibitem{coment_networks} We consider networks of 
nodes possessing equal dynamics connected simultaneously by linear and
nonlinear means. The connecting Laplacian matrix that describes the
topology under which the nodes are connected linearly and nonlinearly
are denoted by ${\mathcal{G}}$ and ${{\mathcal{C}}}$,
respectivelly. The strengths of the linear and nonlinear couplings are
$\sigma$ and $g$ respectivelly. If the nodes in the network are
connected by only linear couplings ($g$=0), ${\mathcal{G}}$ can be an
arbitrary Laplacian matrix. If nodes are simultaneously connected by
linear and nonlinear couplings then ${\mathcal{G}}$ and
${{\mathcal{C}}}$ must commute and every node must receive the same
number $k$ of nonlinear connections comming from other nodes.
\bibitem{comentario} According to the Ruelle formula, for ergodic differentiable systems on compact spaces,
the Kolmogorov-Sinai entropy  is bounded above by the sum of the positive Lyapunov
exponents of the system. If the systems admits an SRB measure, then the Kolmogorov-Sinai
entropy is exactly equal to the sum of the positive Lyapunov exponents of the system
\cite{pesin,young}. For the networks here considered formed by
dissipative systems that possess an attractor whose measure is
completely supported by an unstable manifold, such an equality should
be satisfied. 
In any case, if it is not certain that such an equality holds, notice
that the sum of the positive Lyapunov exponents will be always a
measure of entropy production per unit time, since it measures the
ratio with which partitions should be created in order to define
proper states in a dynamical system.
\bibitem{comentario_LE} While the Lyapunov exponents are obtained
for typical initial conditions (it does not exist two nodes with the
same set of initial conditions), the conditional Lyapunov exponents
are obtained considering that every node has equal initial conditions.
\bibitem{comment_hussein} It is interesting to observe here that this widely
employed synaptic chemical coupling function can be written as
$\Gamma(x_j)=1-F(x_j)$, with $F(x_j)=1/(1+\exp{[\theta(x_j -
\Theta_{syn})]})$. In this way, one may interpret the term
$F(x_j)$ as a Fermi distribution with $1/\theta$ acting as a
temperature and $\Theta$ as the chemical potential. Such a
distribution is a commonality in quantum statistics of Fermion
particles obeying the exclusion principal: no more than one
particle (here a neuron) can occupy the same state.
\bibitem{comenta_conjectura} Our conjecture applies to networks for
  which for every chaotic attractor that possesses a basin of
  attraction with a positive measure there exists also an unstable
  chaotic saddle (the synchronous solution) associated to this chaotic
  attractor. In the case there exists multiple chaotic
  attractors, the coupling strengths $\sigma^*(N)$ and $g^*(N)$ (as well as 
$\sigma_m(N)$ and $g_m(N)$) would
  be a function of the chosen basin of attraction.  The requirement we
  make is that any initial
  condition belonging to an open neighborhood around the unstable chaotic saddle
  goes to only one chaotic attractor. These initial
  conditions are regarded as the typical ones. Therefore, in our
  simulations we consider initial conditions that are small
  perturbations around the unstable synchronous solution.
  Nevertheless, most of the attractors obtained are completely
  out of synchrony. The identification of possible many coexisting
  attractors is a technical matter. In a very general situation, there
  will be hopefully only a few coexisting attractors and one can
  clearly identify the functions $\sigma^*(N)$ and $g^*(N)$ (as well as 
$\sigma_m(N)$ and $g_m(N)$).
\end{thebibliography}
 \end{document}